\documentstyle[11pt,psfig,paspconf]{article}

\begin{document}

\title{Using Perturbative Least Action to Run N Body Simulations Back
in Time}
\author{D. M. Goldberg \& D. N. Spergel}
\affil{Dept. of Astrophysical Sciences, Princeton University,
Princeton, NJ 08544-1001}

\begin{abstract}
In this report, we present a new method for reconstructing N body
initial conditions from a proscribed final density field.  This
method, Perturbative Least Action (PLA) is similar to traditional
least action approaches, except that orbits of particles are found as
expansions around previously determined and physically motivated
orbits.
\end{abstract}


\section{Introduction}
In order to understand the current dynamics and history of large scale
structure, it would be helpful if we were able to generate plausible
initial conditions which would produce structure consistent with
observations.  One of the great difficulties with this problem is that
it is fundamentally ill-posed, since much of the structure of interest
is highly nonlinear.  Many different initial conditions can give rise
to virtually indistinguishable final density fields, even if all of
them obey the constraints given by the Zel'dovich approximation.

We present the idea that even highly nonlinear $N$ body simulations
may be self-consistently run backwards in time.  While previous
attempts at this problem, such as those by Peebles (1989, 1994), Shaya
et al. (1995,1999), and Giavalsco et al. (1993), have suggested using
the least action variational principle to solve the orbits of many
particles, this approach yields the unfortunate result that only one
of the potential solutions (namely, the first infall solution) can be
recovered.  We suggest the novel approach that by using least action
as a perturbation from a known set of (randomly generated) orbits, a
unique solution may be found for that set of initial conditions.  By
performing this analysis with many randomly selected sets of initial
conditions, many sets of self-consistent solutions may be found. In
this way, we may generate realistic initial conditions for interesting
observed structure.

\section{Method}
The least action variational principle states that given a initial and
final positions, a set of particles acting under mutual gravitation
will take the path which minimizes the action (Peebles 1989):
\begin{equation}
S\equiv \sum_{i}\int_0^{t_{0}} dt \frac{a^{2}\dot{\bf x}^2_{i}}{2}-
\frac{\phi_{i}}{2}
\end{equation}
where ${\bf x}_{i}$ is the position of the $i^{th}$ particle given in
comoving coordinates, and $\phi_{i}$ is the potential on the $i^{th}$
particle produced by all the others.

However, let us say that we know that some path ${\bf x}_{i}^{(0)}(t)$
minimizes the action (e.g. the output from an $N$ body code).  In that
case, we may imagine another path:
\begin{equation}
{\bf x}_{i}(t)={\bf x}_i^{(0)}(t)+{\bf x}_i^{(1)}(t)\ ,
\end{equation}
where ${\bf x}_{i}^{(1)}(t_i)$ gives the change in the initial density
field.  Since we know that the action is minimized for the original
path, we may write down the action for this new path as:
\begin{equation}
S=S^{(0)}+\sum_{i}\int_{0}^{t_{0}} dt \left(
a^2\dot{\bf x}_{i}^{(0)}\cdot\dot{\bf x}_{i}^{(1)}+
\frac{a^2 \dot{\bf
x}^{(1)2}_{i}}{2}-\frac{\phi_{i}}{2}+\frac{\phi^{(0)}_{i}}{2} 
\right)
\end{equation}

If we parameterize the perturbation of the paths as:
\begin{equation}
{\bf x}^{(1)}_{i}(t)=D(t){\bf x}^{(1)}_{i}(t_{0})+
\sum_{n}C_{in}^{\alpha}f_n(t)\ ,
\end{equation}
where $D(t)$ is the growth factor of perturbations as given in linear
theory, $\alpha=\{1,2,3\}$ is direction of the vector, and $f_n(t)$
are a set of basis functions, then the perturbed action (and hence the
total action), can be minimized when
\begin{equation}
\frac{\partial S^{(1)}}{\partial C_{in}^{\alpha}}=\int dt
\left[
\dot{f}_{n}(t)a^{2}\dot{\bf x}^{(1)}+f_{n}\left(
\frac{\partial \phi^{(0)}_i}{\partial x^{\alpha}}-
\frac{\partial \phi_{i}}{\partial x^{\alpha}}
\right)
\right]=0 \ .
\label{eq:LA}
\end{equation}
Note that the derivative of the first term in equation (3) is
equivalent to the middle term in equation (5) by virtue of the fact
that ${\bf x}^{(0)}$ is required to minimize action.

Using these equations, a randomly generated particle field (e.g. a
Gaussian random field from some given power spectrum) may be iterated
toward some target density field in the following way.  

First, a set of initial conditions are created using the Zel'dovich
approximation and known power spectrum.  Next, the particle positions
and velocities are evolved and recorded using some Poisson solver.
For our simulations, we have used a straight Particle Mesh (PM) code.
We then compare the final particle field to the target density field.
Next, we determine perturbations (values of ${\bf
x}_{i}^{(1)}(t_{0})$) which would cause the final density field to
more strongly match the target field.  Using those perturbations, we
find the values of $C_{in}^{\alpha}$ which solve
equation(~\ref{eq:LA}).  Using those coefficients, we determine the
change in positions (and hence density) of the particles at high
redshift, which gives us a new set of initial conditions.  This new
set of initial conditions can be run through the $N$ body code, and
the process may be iterated until a satisfactory fit is returned.

While the perturbed initial field is not, strictly speaking, Gaussian
random, it does obey the Zel'dovich approximation by an appropriate
selection of basis functions.  Moreover, since the unperturbed initial
field was Gaussian random, and since the perturbative least action
approach aims to find the closest set of initial conditions to those
randomly generated (as illustrated in Figure 1) which will satisfy the
proscribed final conditions, the solution will be fairly close to
Gaussian random and as close as possible to the given power spectrum.

\section{Simulations}

As a test of this scheme, we create a highly nonlinear target density
field with three overlapping isothermal spheres, with peaks as high as
$\delta=5$.  We then take two different sets of initial density fields
(realizations of the same power spectrum), and iterate using the
Perturbative Least Action (PLA) principle.  We use a Particle Mesh
code as our Poisson solver.

The simulations are each $64^3$ gridcells and $32^3$ particles, and
were run from $z=99$ to $z=0$.  Twenty timeslices or positions and
velocities are used in order to do the least action integration.
Around six iterations (i.e. computation of the least action, and
running the result through the PM code) were necessary to produce the
results shown.
 
\begin{figure}[hp]
\centerline{\psfig{figure=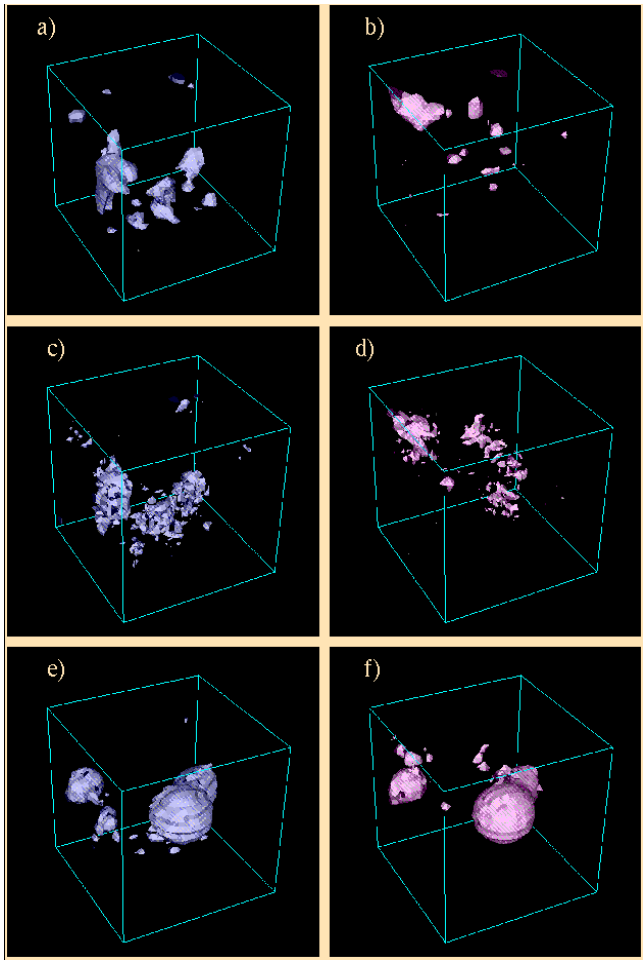,height=5.0in,angle=0}}
\caption{The results of a set of simulations.  Panels a-b show two
random realizations of the initial power spectrum.  Panels c-d show
the perturbed initial fields (after solving using least action).
Panels e-f show the perturbed fields after having been run through the
PM code.}
\end{figure}

Figure 1 shows the results of these simulations.  The top row of
panels show two randomly generated density fields at z=99.  The
velocity fields in each are given by the Zel'dovich approximation.  By
applying perturbative least action to each of these sets of initial
fields with a particular target final density field, a new set of
``perturbed'' initial fields may be created.

The second row shows the perturbed initial fields.  Notice that the
large scale perturbations remain unchanged, and that only the small
scale perturbations seem affected.  This is due to the fact that we
have specified the target field on a cell by cell basis, necessitating
a very large amount of small scale power.  Finally, the bottom panels
show the result of integrating from our perturbed initial conditions.
Despite the widely different initial conditions, both final fields
strongly resemble both each other and the target field.

Though this toy problem is presented as proof of method, it is clear
that this principle is applicable to a number of more complex problems
such as determination of small scale primordial power, cluster
redshift surveys, and study of the Local Group.  This last would be
quite interesting as recent studies (see e.g. Mateo 1998, and
references therein) give a rather detailed picture of the current
local density field and a series of simulations which could provide
insight into probable infall scenarios would be most illuminating.

\acknowledgments
We would like to thank P.J.E. Peebles, Michael Strauss, Jerry Ostriker
and Michael Vogeley, for helpful suggestions, and Michael Blanton for
his invaluable visualization software.  DMG is supported by an NSF
graduate research fellowship.

\end{document}